\documentclass{jnmp}

\usepackage{amsmath}
\usepackage{epsfig}

\JNMPnumberwithin{equation}{section}

\begin{document}

\renewcommand{\evenhead}{P~Bracken and A~M~Grundland}
\renewcommand{\oddhead}{On Complete Integrability of the
 Generalized Weierstrass System}

\thispagestyle{empty}

\FirstPageHead{9}{2}{2002}{\pageref{bracken-firstpage}--\pageref{bracken-lastpage}}{Article}

\copyrightnote{2002}{P~Bracken and A~M~Grundland}

\Name{On Complete Integrability\\ of the
 Generalized Weierstrass System}
\label{bracken-firstpage}

\Author{P~BRACKEN and A~M~GRUNDLAND$^{\, *}$}

\Address{Centre de Recherches Math\'{e}matiques, Universit\'{e} de Montr\'{e}al, \\
2920 Chemin de la Tour,  Pavillon Andr\'{e} Aisenstadt,  C. P. 6128 Succ. Centre Ville,   \\
Montr\'{e}al, QC, H3C 3J7 Canada   \\
$^{*}$Department of Mathematics, Universit\'e du Qu\'ebec, Trois-Rivieres, QC, A9G 5H7 Canada \\
E-mail: bracken@crm.umontreal.ca, \ grundlan@crm.umontreal.ca}

\Date{Received November 18, 2001; Revised  December 7, 2001;
Accepted January 1, 2002}

\begin{abstract}
\noindent In this paper we study certain aspects of the complete
integrability of the Genera\-lized Weierstrass system in the
context of the Sinh-Gordon type equation. Using the conditional
symmetry approach, we construct the B\"{a}cklund transformation
for the Generalized Weierstrass system which is determined by
coupled Riccati equations. Next a linear spectral problem is found
which is determined by nonsingular $2 \times 2$ matrices based on
an $sl (2, \mathbb C)$ representation. We derive the explicit form
of the Darboux transformation for the Weierstrass system. New
classes of multisoliton solutions of the Generalized Weierstrass
system are obtained through the use of the B\"{a}cklund
transformation and some physical applications of these results in
the area of classical string theory are presented.
\end{abstract}

\section{Introduction}

The theory of surfaces has recently had a large resurgence
of interest in many applications to very different and diverse
areas of theoretical physics. This is in no small measure
due to the fact that many equations which are of interest
in various areas of mathematical physics also appear, or
can be incorporated, in the study of surfaces in three-dimensional
space. In particular, the Sine and Sinh-Gordon equations
are of a great significance in this respect. Moreover, the theory
of constant mean curvature surfaces has been of continued
importance in the study of many problems which have physical
applications. For example, minimal surfaces and constant mean curvature
surfaces in particular have recently found applications in
the areas of two-dimensional gravity~[1], quantum field theory
as well as in the area of string theory~[1]. As a string
propagates through space-time, it describes a surface in
space-time which is called its world sheet. The study of
classical strings in three and four-dimensional space is a crucial
first step in producing a quantum theory of strings.
It can be said that first quantized string theory is the
study of conformal field theories on Riemann surfaces.

The stationary two-dimensional sigma model has been shown to be
of use in generating two-dimensional surfaces
immersed in three-space, and there
are links between this model and other models,
such as the non-Abelian Chern--Simons theories, which are
of great importance in certain areas of condensed matter
physics~[1]. It has been shown that the Chern--Simons
gauged  Landau-Ginsburg model plays the role of effective
theory for the Fractional Quantum Hall Effect~[2].
The Chern--Simons equation of motion can describe time evolving
two-dimensional surfaces in such a way that the deformation is
not only locally compatible with the Gauss--Codazzi equation,
but completely integrable as well~[3].
In the static
limit, the self-dual version of the model possesses soliton
solutions. These correspond to Laughlin's quasiparticles
and give a realization of anyon quasiparticles. On the
other hand, the self-dual Chern--Simons model can be
associated with the stationary two-dimensional continuous
classical Heisenberg model, which can be related to the
two-dimensional sigma-model.

Another area of recent interest with regard to applications is to
the area of liquid crystals and the theory of membranes~[4]. Fluid
membranes may be idealized as two-dimensional surfaces with each
membrane being made up of a double layer of long mo\-le\-cu\-les.
Various physical properties of interest such as elastic free
energy per unit area can be calculated in terms of quantities
which are directly related to the geometry of the surface. In fact
the curvature elastic free energy per unit area of the membrane
can also be formulated rigorously in terms of two-dimensional
differential invariants of the surface. It is of considerable
interest with regard to these types of physical applications to
obtain shape equations for the membrane. These interrelate the
basic parameters and functions which determine the form of a given
membrane surface, as in a liquid crystal. In an equilibrium state,
the energy of any physical system must be minimized. One usually
writes down a~shape energy function~$F$ in terms of the basic
parameters and then minimizes it, and the result is a~shape
equation. An example of such a shape function is given by
\[
F = \frac{1}{2} k_{c} \int (2H + c_{0})^2  dA + \Delta p
\int dV + \lambda \int  dA,
\]
where $k_{c}$ is the bending rigidity of the membrane, $H$ the
mean curvature and the spontaneous curvature $c_{0}$ takes
account of the asymmetry effect of the membrane or the surrounding
environment. The pressure difference between the outside and the inside
of the membrane is called $\Delta p$, $\lambda$ the tensile
stress acting on the membrane. Mathematically, $\Delta p$ and
$\lambda$ may be considered as Lagrange multipliers. The shape equation
is obtained from the first variation of this $F$.
Specific Delaunay's surfaces of constant mean curvature can be
written as $\sin \psi (\rho) = a \rho
+ d /\rho$. This equation is then substituted
into a given shape equation and results in constraint
equations between the parameters $a$ and $b$, which
give a characterization of the surface~[4, p.~114].

In this paper, we study a connection between the
Generalized Weierstrass (GW) system
inducing constant mean curvature surfaces immersed in $\mathbb R^3$ and a
Sinh-Gordon type equation. The objective of this paper is, using this
link between these
two systems, to derive a representation of a linear
spectral problem for which the matrices are nonsingular,
and find the corresponding Darboux and B\"{a}cklund
transformations for the GW system.
These transformations are derived here for the first time.
Based on these transformations,
we construct solutions of the GW system
and investigate minimal surfaces immersed in three-dimensional
Euclidean space.

This paper is organized as follows. In Section~2,
a short presentation of the conditional symmetry approach for
partial differential equations which admit the Painlev\'e
property is given. In Section~3, using the conditional symmetries
we derive the linear spectral problem, and we find the Darboux
and B\"{a}cklund transformations for the GW system.
Section~4 contains new examples of
multi-soliton solutions of the GW system and some physical
interpretations of these results are given
in the area of classical string theory.

\section{Conditional symmetries}

In this section, we give a brief overview of the conditional
symmetry approach for PDEs as developed in~[5--8]. In this
context, we concentrate on examining certain aspects of
integrability of $k$-th order nonlinear PDEs. The technique which
is outlined below is applied only to such classes of PDE which
pass the Painlev\'{e} test and can be presented in the form of a
polynomial in the unknown variable $u$ and its derivatives,
possibly after a transformation in the space of independent and
dependent variables $X \times U$. The basic terminology and
notation used here in the application of Lie groups to
differential equations are in conformity with~[9]. We are
particularly interested in combining singularity structure
analysis and Lie point symmetries in order to recover the
Auto-B\"{a}cklund transformation (Auto-BT) and Darboux
transformation (DT) for PDEs if such exist. In the literature,
several attempts at treating this subject can be found, for
example~[5--8], and references therein. More recently, this
subject has been studied for first order systems of PDEs, by one
of the authors, leading to the development of a new version of the
conditional symmetry method~[10,~11]. This approach, like other
nonclassical methods (see for a review of the subject~[12]) makes
possible the explicit determination of certain classes of
solutions, invariant under a group of transformations which maps a
subset of solutions of the initial equation into other solutions
of a different equation, that is, a subsystem composed of the
initial PDE and differential constraints (DCs), which are mutually
consistent. In this presentation, we postulate in accordance with
the method worked out in~[11], that these multiple DCs take a
specific form for which all derivatives of the unknown
function~$u$ are expressible in terms of some functions of the
independent and dependent variables only. Hence we consider the
overdetermined system composed of a nondegenerate $k$-th order
scalar PDE and a first order system of DCs
\begin{gather}
  \Delta \left(x, u^{(k)}\right) = 0,\\
 Q_{i} \left(x, u^{(k)}\right) \equiv \frac{\partial u}{\partial x^{i}} - \phi_{i} \left(x,u^{(k)}\right)=0,
\qquad i=1, \ldots, p
\end{gather}
in $p$ independent variables $x = \left(x^{1}, \ldots , x^{p}\right)$ which form
some local coordinates in Euclidean space $X$. The compatibility
conditions for (2.1) and (2.2) are given by
\begin{gather}
~(i) \quad  \phi_{[i,j]} + \phi_{[j} \phi_{i],u} = 0,  \qquad
i,j = 1, \ldots, p,\nonumber  \\
(ii)  \quad  \Delta\left(x,u,\phi^{(k-1)}\right) = 0,  \qquad
\phi = (\phi_{1}, \ldots , \phi_{p}).
\end{gather}
The brackets $[i,j]$ denote the alternation with respect to the
indices $i$ and $j$, that is,
\begin{gather*}
\phi_{[i,j]} \equiv 2( \phi_{i,j} - \phi_{j,i}),\\
\phi_{[i} \phi_{j],u} = 2(\phi_{i} \phi_{j,u} - \phi_{j} \phi_{i,u}).
\end{gather*}
Note the equation (2.2) means that the characteristics $Q_{i}$
of a set of $p$-linearly independent vector fields
(defined on $X \times U$)
\begin{equation}
Z_{i} = \partial_{x^{i}} + \phi_{i} (x,u) \partial_{u}, \qquad
i= 1, \ldots, p,
\end{equation}
are equal to zero.

An Abelian Lie algebra $L$ spanned by the vector fields  $Z_{1},
\ldots, Z_{p}$ is called a conditional symmetry algebra of the
$k$-th order PDE (2.1), if the vector fields $Z_{1}, \ldots,
Z_{p}$ are tangent to the subvariety
\[
S = S_{\Delta} \cap S_{Q},
\]
where we associate the initial system $\Delta : J^{k} \rightarrow \mathbb R$
and a first order system of DCs
$Q_{i} : J^{1} \rightarrow \mathbb R^{p}$ with the subvarieties of the
solution spaces
\begin{gather*}
S_{\Delta} = \left\{ \left(x, u^{(k)}\right) \in J^{k} : \Delta\left(x, u^{(k)}\right)=0 \right\},
\\
S_{Q} = \left\{ \left(x, u^{(1)}\right) \in J^{1} : Q_{i} \left(x, u^{(1)}\right) = 0, i=1, \ldots,
p\right\},
\end{gather*}
respectively.
This definition means that the $k$-th prolongation of the vector fields
$Z_{i}$ belongs to the tangent space to $S$ at $\left(x, u^{(k)}\right)$, that is,
\begin{equation}
{\rm pr}^{(k)}  Z_{i} |_{S} \in T_{\left(x, u^{(k)}\right)} S, \qquad
i=1, \ldots, p.
\end{equation}

A solution $u=f(x)$ of the $k$-th order PDE (2.1) is called a conditionally
invariant solution, if its graph $\{ (x, f(x)) \}$ is invariant under
an Abelian distribution of the vector fields $Z_{1}, \ldots, Z_{p}$
satisfying conditions~(2.5).

It has been shown [11] that a nondegenerate $k$-th order PDE (2.1)
admits a $p$-di\-men\-sio\-nal conditional symmetry algebra $L$
if and only if there exists
a set of $p$ linearly independent vector fields (2.4) for which the
$C^{k-1}$ functions $\phi_{i}$ satisfy the conditions (2.3). The graph
of a solution of the overdetermined system
composed of (2.1) and (2.2) is invariant under
the vector fields $Z_{i}$, $1 \leq i \leq p$. Hence according to the
above definition, this means that there exists a conditionally
invariant solution of PDE (2.1).

It has been proved [13] that any PDE (2.1) of the $k$-th order
admits infinitely many compatible first order DCs. However, this
statement shows only existence of such constraints, but does not
provide a constructive method for finding the explicit form of
these DCs. Thus, the construction of conditional symmetries is
reduced to the selection of such subsystems composed of initial
PDE (2.1) and DCs (2.2) for which conditions (2.3) hold. In
general, system (2.3) is a nonlinear one and usually very
difficult to solve, except in some particular cases. Nevertheless,
there exist many physically interesting systems of PDEs for which
particular solutions of (2.3) lead to B\"{a}cklund transformations
described by first order differential constraints or to solutions
depending on some arbitrary constants~[14]. These particular
solutions of (2.3) are obtained by expanding each
function~$\phi_{i}$ into a polynomial in the dependent
variables~$u$. This polynomial is reduced often to a~trinomial.
This means that equations (2.2) become a Riccati system of
equations which possess the Painlev\'e property.

In Section~3, we show on a specific example of the
generalized Weierstrass system
inducing constant mean curvature surfaces in $\mathbb R^3$,
that the conditions (2.3) for the existence of
conditional symmetries can be used to construct a certain class
of B\"{a}cklund and Darboux transformations.
This construction consists
of the following steps.

First, we assume that PDE (2.1) passes the Painlev\'{e} test. This
means that (2.1) satisfies the necessary conditions for the absence of
movable critical singularities in its general solution on arbitrary
noncharacteristic surfaces~[15,~16]. This test provides us with a
tool for assessing the integrability of PDEs. At least we can reject
as nonintegrable these PDEs which do not pass the Painlev\'e test.

We restrict our considerations to the particular case when the
singularity structure of a~solution of PDE~(2.1) consists of only poles.
According to Painlev\'e analysis for PDEs~[16], it has been shown that
if the choice of the expansion variable $\chi$ in
terms of the singular manifold
variable $\varphi (x) = \varphi_{0}$, (where $\chi$
vanishes as $\varphi - \varphi_{0}$), then the beginning of the
Laurent series for the solution $u$ takes place at the order $-n$
in $\chi$, that is,
\begin{equation}
u  \sim  \chi^{-n},  \qquad  n \in \mathbb Z <0,
\end{equation}
where the power $(-n)$ denotes the multiplicity of a pole.

Secondly, we postulate that the difference of two distinct solutions $u$
and $\hat{u}$ of PDE~(2.1) can be represented in a polynomial form
in terms of an auxillary variable $y$ up to the degree~$(-n)$.
The variable $y$ is
a mapping of space $X$ into some $m$-dimensional space~$B$. This
demand implies that we perform an embedding transformation of
the variable $(u - \hat{u})$ into the space $B$ with coordinates
$y = \left(y^{1}, \ldots, y^{m}\right)$. Thus we assume that it can be realized through
a specific Darboux transformation of the form
\begin{equation}
u - \hat{u} = \sum_{J} c_{J} y^{J},  \qquad
1 \leq \#J \leq (-n),
\end{equation}
where $J= \left(j_{1}, \ldots , j_{m}\right)$ and $j_{i}$ is a positive integer
such that
\begin{equation}
\#J = j_{1} + \cdots + j_{m} = -n.
\end{equation}
The coefficients $c_{J}$ in the expansion (2.7) are assumed to be constants.
Eliminating the function $u$ in the initial equation (2.1) through
transformation (2.7), we obtain a $k$-th order PDE for the
unknown function $y$. Denote this PDE by
\begin{equation}
\Delta_{1} \left(x, \hat{u}^{(k)}, y^{(k)}\right) = 0,
\end{equation}
where $\hat{u}^{(k)}$ is a given function of $x$. Equation (2.9)
is the starting point for our further analysis.
In this section, we focus on a close connection between the conditional
symmetries and B\"{a}cklund transformations associated with nonlinear
systems of PDEs (2.9). This connection is mainly due to the fact
that the set of DCs (2.2) in the variable $y$ admits a~superposition
formula (SF), as is also the case for BTs~[17]. Based on the
application of Lie's theorem on fundamental sets of solutions~[18]
to the case of first order PDEs~[17],
we show this link on a specific example in the
next section.  We assume that these DCs take the
particular form of coupled matrix Riccati equations based on
some given representation of the Lie algebra. This means that the DCs
(2.2) take a specific form for which the first derivatives of $y$
are decomposable in terms of $x$ and $y$ as follows
\begin{equation}
\frac{\partial y^{\alpha}}{\partial x^{i}} (x) = \sum_{l=1}^{r}
A_{i}^{l} (x) b_{l}^{\alpha} (y(x)).
\end{equation}
The set of functions $b_{l}^{\alpha}$ is identified with
linearly independent vector fields on a space $B$
\begin{equation}
\hat{b}_{l} = b_{l}^{\alpha} (y) \partial_{y_{\alpha}}, \qquad
l=1, \ldots r,
\end{equation}
which generate a finite-dimensional Lie algebra ${\cal G}$
\begin{equation}
[ \hat{b}_{l}, \hat{b}_{k} ] = C_{lk}^{a} \hat{b}_{a},
\qquad
1 \leq a, k, l \leq r,
\end{equation}
where $C^{a}_{lk}$ are constants of the assumed Lie algebra structure.
Note that the problem of construction and classification
of all finite-dimensional Lie algebras which can be realized in
terms of operators (2.11) remains an open one~[18--20]. Nevertheless,
this subject has been recently extensively investigated and there exist
numerous lists of finite-dimensional Lie algebras which can
serve as a source for our selection (see for example~[19,~20] and
references therein).

We select one of the finite-dimensional Lie algebras ${\cal G}$
and its representation in terms of vector fields (2.11) with
polynomial coefficients in the variable $y$. We start with the
lowest-dimensional algebra and, if necessary, proceed to consider
the higher dimensional ones. For a chosen Lie algebra ${\cal G}$
the right hand side of expression (2.10) becomes a~polynomial in
the dependent variable $y$. Substituting (2.10) repeatedly $(k-1)$ times into
PDE (2.9) and next requiring that the coefficients of successive
powers of $y$ in the equation so obtained vanish, we get a system
of $(k-1)$ order PDEs for the functions $A_{i}^{l}$. Denote
this system by
\begin{equation}
Q^{a} \left(x, \hat{u}^{(k)} (x) , A_{i}^{l(k-1)}\right) =0,
\qquad a =1, \ldots , r.
\end{equation}
For the assumed Lie algebra ${\cal G}$ the compatibility
conditions for DCs (2.10) impose zero gauge curvature conditions
on the functions $A_{i}^{l}$
\begin{equation}
A^{a}_{[i,j]} + \frac{1}{2} C_{lk}^{a} A_{i}^{l} A_{j}^{k} =0,
\qquad i \neq j = 1, \ldots,p.
\end{equation}
As a result, we arrive at an overdetermined system of equations
for the functions $A_{i}^{l}$. We denote this system, consisting
of equations (2.13) and (2.14), by $H$. To establish the existence
of solutions of this overdetermined system, an analysis of the
compatibility conditions of the system is required.

It has been shown~[21] on several examples, such as the
AKNS, Boussineq, Kadomtsev--Petviashvili, Sawada--Kotera, and
Tzitzeica equations, that there exist
many physically interesting systems of PDEs for which the
general solution of the system $H$ lead to Auto-B\"{a}cklund
transformations. This phenomena takes place when the set of solutions
of the system $H$ is parametrized by a function $\hat{u}$
satisfying the original PDE (2.1) and by at least one constant
parameter $\lambda$. In this case, the functions $A_{i}^{l}$ can
be expressed uniquely in terms of the old solution $\hat{u}$ of
(2.1) and some constant $\lambda$. Then DCs (2.10) become
\begin{equation}
\frac{\partial y^{\alpha}}{\partial x^{i}} (x) =
\sum_{l=1}^{r} A_{i}^{l} (\hat{u}, \lambda)
b_{l}^{\alpha} (y(x)).
\end{equation}

From each solution $\hat{u}$ of PDE (2.1) we integrate DCs (2.15)
for the functions $y^{\alpha}$ and find a solution $u$ of
original equation (2.1) via Darboux transformation~(2.7). Thus,
system~(2.15) together with~(2.7) determines a specific Auto-BT between
sets of solutions of the initial PDE~(2.1). By eliminating the auxilliary
variables $y^{\alpha}$ from the system~(2.15) and~(2.7) we can obtain
an explicit form of this Auto-BT.

Note that the solution of (2.9) which is obtained from the proposed procedure
constitutes a conditionally invariant solution since it is
invariant under the $p$-dimensional conditional symmetry algebra $L$
spanned by the vector fields of the form
\begin{equation}
Z_{i} = \partial_{x^{i}} +
\sum_{l=1}^{r} A_{i}^{l} ( \hat{u}, \lambda) \hat{b}_{l}, \qquad
i=1, \ldots, p,
\end{equation}
where $\hat{b}_{l} = b_{l}^{\alpha} (y) \partial_{y^{\alpha}}$
generate a finite-dimensional Lie algebra (2.12).

\section{The generalized Weierstrass system}

We now proceed to apply the conditional symmetry approach to the case of the
generalized Weierstrass system (GW) inducing constant mean
curvature surfaces embedded
in $\mathbb R^{3}$ as derived by B~Konopelchenko in~[22].
This system is described by a set of Dirac type equations
for two complex fields $\psi_{1}$ and $\psi_{2}$ given by
\begin{gather}
\partial \psi_{1} = p \psi_{2}, \qquad   \bar{\partial} \psi_{2} =-p \psi_{1}, \nonumber\\
\bar{\partial} \bar{\psi}_{1} = p \bar{\psi}_{2}, \qquad
\partial \bar{\psi}_{2} =-p \bar{\psi}_{1},
\end{gather}
where $p = |\psi_{1}|^{2} + |\psi_{2}|^{2}$,
$\partial = \partial/ \partial z$ and $\bar{\partial} = \partial/
\partial \bar{z}$.

Equations (3.1) possess several conserved quantities
\begin{gather}
\bar{\partial} \left(\bar{\psi}_{1} \partial \psi_{2} - \psi_{2}
\partial \bar{\psi}_{1} \right) = 0,  \qquad
\partial \left( \psi_{1} \bar{\partial} \bar{\psi}_{2} - \bar{\psi}_{2}
\bar{\partial} \psi_{1} \right) = 0,   \nonumber\\
\partial (\psi_{1})^{2} + \bar{\partial} (\psi_{2})^{2} = 0,  \qquad\quad \ \,
\bar{\partial} \left( \bar{\psi}_{1}\right)^{2} + \partial \left(\bar{\psi}_{2}\right)^{2} = 0.
\end{gather}
It has been shown [22] as a consequence of these conservation laws that
there exist three real valued functions $X^{i} (z, \bar{z})$, $i=1,2,3$
such that
\begin{gather}
X_{1} + i X_{2} = 2i \int_{\Gamma} \left( \bar{\psi}_{1}^{2} \,dz' -
\bar{\psi}_{2}^{2} \,d \bar{z}'\right),    \qquad
X_{1} - i X_{2} = 2i \int_{\Gamma} \left( \psi_{2}^{2} \, dz' -
\psi_{1}^{2} \, d \bar{z}'\right),    \nonumber\\
X_{3} = -2 \int_{\Gamma} \left( \bar{\psi}_{1} \psi_{2} \, dz' +
\psi_{1} \bar{\psi}_{2} \, d \bar{z}'\right),
\end{gather}
Note that by virtue of the conservation laws (3.2), the
right hand sides in expression~(3.3) do not depend on the
choice of contour $\Gamma$ in the complex plane $\mathbb C$,
but only on its endpoints.
This is due to the fact that the
integrals (3.3) have the form
\[
\int_{\Gamma} F(z, \bar{z}) \, dz + \bar{F} (z, \bar{z}) \, d \bar{z},
\]
which satisfy the condition
\[
\bar{\partial} F = \partial \bar{F}.
\]
Consequently, the differentials of equations (3.3) are exact ones.
The functions
$X^{i} (z, \bar{z})$ can be identified as the coordinates of the position
vector $\vec{X}$ of a surface embedded in $\mathbb R^{3}$. Hence from
equations (3.1) and (3.3), we can determine the induced metric
of this surface
\begin{gather}
ds^{2} = 4 p^{2} \, dz \, d \bar{z},
\end{gather}
in isothermic coordinates.
The Gaussian curvature and constant mean curvature can be
evaluated from
$K=- p^{-2} \partial \bar{\partial} ( \ln p)$,
$H=1$, respectively, where $p$ is given in (3.1).

The system (3.1) is known to be completely integrable, and a Lax pair
for it has been found recently by authors~[23,~24].
However, the matrix appearing in the Lax pair is singular and
nilpotent, which prevents the construction of solutions of~(3.1). The objective
of this paper is to demonstrate that it is possible
to find such a representation of the Lax pair for which the
matrices are nonsingular. This representation is suitable for
constructing solutions of GW system (3.1). Next, we construct the
B\"{a}cklund transformation and the Darboux transformation by
making use of a close connection between these transformations and
conditional symmetries.
Finally, based on these transformations, we construct several new
classes of multisoliton solutions for (3.1)
and investigate the corresponding surfaces in $\mathbb R^3$.

In our investigation of the integrability of the GW system (3.1)
we change the dependent variables $\psi_{1}$ and $\psi_{2}$ to
new dependent variables
\[
p = |\psi_{1}|^{2} + |\psi_{2}|^{2},
\]
and the current
\begin{equation}
J = \psi_{1} \bar{\partial} \bar{\psi}_{2} - \bar{\psi}_{2}
\bar{\partial} \psi_{1},
\end{equation}
in order to simplify its structure. In terms of these new variables,
we show that GW system (3.1) can be decoupled into a direct
sum of the elliptic Sh-Gordon type equation and the conservation
of current $\bar{\partial} J=0$.

In fact, differentiating the function $p$
with respect to $z$ and $\bar{z}$, we obtain,
\begin{gather}
\partial p = \psi_{1} \left(\partial \bar{\psi}_{1}\right) + \bar{\psi}_{2}
\left(\partial \psi_{2}\right), \qquad  \bar{\partial} p =
\bar{\psi}_{1} \left( \bar{\partial} \psi_{1}\right) +
\psi_{2} \left( \bar{\partial} \bar{\psi}_{2}\right),  \nonumber \\
\partial \bar{\partial} p = \bar{\partial} \psi_{1}
\partial \bar{\psi}_{1} + \bar{\partial} \psi_{2} \partial \psi_{2}
- p^{3}.
\end{gather}
Making use of conservation laws (3.2) and equations (3.6), GW
system (3.1) takes the decoupled form in the variables $p$ and $J$,
\begin{gather}
\partial \bar{\partial} \ln p = \frac{|J|^{2}}{p^{2}} - p^{2},
\qquad  \bar{\partial} J = 0,
\qquad  \partial \bar{J} =0.
\end{gather}
If we introduce the new dependent variable
\[
p = e^{\varphi/2},
\]
into equation (3.7), we then obtain an elliptic sinh-Gordon type
equation of the form~[25]
\begin{gather}
\partial \bar{\partial} \varphi =-4 \sinh \varphi -2
\left(1 - |J|^{2}\right) e^{-\varphi},
\qquad
\bar{\partial} J = 0.
\end{gather}
In particular, if the modulus of the current $J$ is different
from zero, $|J| \neq 0$, then we can introduce new independent,
dependent variables $\eta$, $\bar{\eta}$ and $\omega$
\[
d \eta = J^{1/2} \, dz,
\qquad
d \bar{\eta} = \bar{J}^{1/2} \, d \bar{z},
\qquad
\omega = \frac{p^{2}}{|J|},
\]
respectively, such that GW system (2.1) takes the decoupled form
\[
( \ln \omega)_{\eta \bar{\eta}} = 2 \left(\frac{1}{\omega} - \omega\right),
\qquad
\bar{\partial} J = 0,
\qquad
\partial \bar{J} = 0.
\]
Hence GW system (2.1) becomes a direct sum of the elliptic
Sh-Gordon equation and the conservation of current $J$.

Equation (3.7) has the Painlev\'e property. Consequently, we obtain
that the general solution $p$ of (3.7) admits double poles with
two residues of opposite sign
\begin{gather}
p^{\pm 2} = e^{ \pm \varphi}  = \pm \chi^{-2},
\end{gather}
where $\chi$ is the expansion variable of the Laurent series.
According to the proposed procedure, based on the approach described
by Conte and Musette~[26]
we assume a specific form of the
Darboux transformation for the case when
equation (3.8) admits opposite residues
\begin{gather}
\varphi - v = 2 \ln \frac{\phi_{1}}{\phi_{2}}.
\end{gather}
The function $v$ satisfies the initial equation (3.8)
and quantities $\phi_{1}$ and $\phi_{2}$ are two entire functions.
Note that a similar situation for double poles with opposite
residues arises in the study of singularity structure for
sine-Gordon and MKDV equations~[26]. Introducing a new
variable $y= \phi_{1}/ \phi_{2}$ and changing the variables
in (3.10) according to
\begin{gather}
p = e^{\varphi/2}, \qquad
q = e^{v/2},
\end{gather}
we find that the Darboux transformation for
equation (3.7) can be realized by the following expression
\begin{gather}
p=q \, y,
\end{gather}
or equivalently,
\[
\varphi = v - 2 \ln y.
\]
A first step on the way to constructing a BT for (3.7) is to look for a
conditional symmetry algebra $L$ spanned by two vector fields
$Z_{1}$, $Z_{2}$ which have the characteristic equations of the form
(2.10). We have to assume a specific Lie algebra structure for the
generators $\{ \hat{b}_{l} \}$. We start the analysis with the
lowest-dimensional case, namely that of the $sl(2, \mathbb C)$ algebra
which admits the one-dimensional representation $(\partial_{y}, y \partial_{y},
y^{2} \partial_{y})$ in terms of a coordinate~$y$.
This algebra comes up in the study of several completely integrable
models eg.~[19].
In our case, DCs (2.10) in one complex variable $y$ take the form
of coupled scalar Riccati equations with nonconstant coefficients
\begin{gather}
\partial y = A_{1}^{0} (z, \bar{z}) + A_{1}^{1} (z,\bar{z}) y
+ A_{1}^{2} (z, \bar{z}) y^{2},  \nonumber\\
      \bar{\partial} y = A_{2}^{0} (z, \bar{z}) + A_{2}^{1} (z, \bar{z}) y
+ A_{2}^{2} (z, \bar{z}) y^{2}.
\end{gather}
The zero curvature conditions for (3.13) are given by
\begin{gather}
\bar{\partial} A^{0}_{1} - \partial A_{2}^{0} + A^{1}_{1} A^{0}_{2}
- A^{0}_{1} A_{2}^{1} = 0,\nonumber\\
\bar{\partial} A_{1}^{1} - \partial A_{2}^{1} + 2
\left(A^{2}_{1} A^{0}_{2} - A^{2}_{2} A^{0}_{1}\right)  = 0,\nonumber\\
\bar{\partial} A^{2}_{1} - \partial A^{2}_{2} -
A^{1}_{1} A^{2}_{2} + A_{1}^{2} A_{2}^{1} = 0.
\end{gather}
The substitution of the new variables (3.11) and the Ansatz
(3.12) into equations (3.7) gives
\begin{gather}
\frac{1}{q y} ( \partial \bar{\partial} q \, y + \partial q \,
\bar{\partial} y + \bar{\partial} q \, \partial y + q
\bar{\partial} \partial y) - \frac{1}{q^{2} y}
( \partial q \, y + q \, \partial y) ( \bar{\partial} q )
\nonumber\\
\qquad {}- \frac{1}{q y^{2}} ( \partial q \, y + q \, \partial y)
\bar{\partial} y - \frac{|J|^{2}}{q^{2} y^{2}} + q^{2} y^{2} = 0.
\end{gather}
Using equations (3.13) we can eliminate the derivatives of the
complex variable $y$ in the expression (3.15). Next, we require
that the coefficients of the successive powers of $y$ in the
equation so obtained vanish, to give the system
\begin{align}
& (1) &\!\!\! & q^{2} A^{0}_{1} A^{0}_{2} + |J|^{2} =0,  &  &  q^{2} A^{0}_{1}
A^{0}_{2} + |J|^{2} = 0,   \nonumber\\
 & (2) &\!\!\!& \partial A^{0}_{2} - A^{1}_{1} A^{0}_{2} = 0, &  &
\bar{\partial} A^{0}_{1} - A^{0}_{1} A^{1}_{2} = 0,   \nonumber\\
 & (3) &\!\!\!& \partial A^{1}_{2} + A^{2}_{2} A^{0}_{1} - A^{2}_{1} A^{0}_{2}
+ \partial \bar{\partial} \ln q = 0,\!\! &   &
\bar{\partial} A^{1}_{1} + A^{2}_{1} A^{0}_{2} - A^{0}_{1} A^{2}_{2}
+ \bar{\partial} \partial \ln q = 0,  \\
& (4) &\!\!\!& \partial A^{2}_{2} + A^{2}_{2} A^{1}_{1} = 0,  &  &
\bar{\partial} A^{2}_{1} + A^{2}_{1} A^{1}_{2} = 0,   \nonumber\\
&(5) &\!\!\!& A^{2}_{2} A^{2}_{1} + q^{2} = 0,  &  &
A^{2}_{1} A^{2}_{2} + q^{2} = 0.   \nonumber
\end{align}
We obtain an overdetermined system (denoted by $H$ in the
previous section) composed of (3.14)
and (3.16) for the unknown functions $A^{k}_{i}$.
In our case, this system is consistent since the compatibility
conditions are identically satisfied.
This system has a nontrivial,
unique solution for the $A_{i}^{k}$. We briefly outline how
this solution can be obtained.
The second and fourth equations in the first column of (3.16)
can be written in the form
\[
\partial \ln A^{0}_{2} = A_{1}^{1},  \qquad
- \partial \ln A^{2}_{2} = A^{1}_{1}.
\]
Equating these equations, we can integrate to obtain
\begin{gather}
A^{0}_{2} = \frac{\bar{g}(\bar{z})}{A^{2}_{2}},
\end{gather}
where $\bar{g}$ is a complex function of $\bar{z}$.
Similarly, from the second and fourth equations in the second
column of (3.16), we obtain
\[
\bar{\partial} \ln A^{0}_{1} = - \bar{\partial} \ln A^{2}_{1},
\]
which can be solved to give,
\begin{gather}
A^{0}_{1} = \frac{h(z)}{A^{2}_{1}},
\end{gather}
where $h$ is a complex function of $z$.
Substituting these results into the first equation in~(3.16),
we find that
\[
h(z) \bar{g} ( \bar{z}) - |J|^{2} = 0.
\]
Thus, without loss of generality, one may take
$h(z) = J(z)$ and $\bar{g}( \bar{z}) = \bar{J} (\bar{z})$.

From the first equation in (3.16), we can write
\begin{gather}
A^{0}_{1} = - \frac{|J|^{2}}{q^{2} A^{0}_{2}},
\end{gather}
thus $A^{0}_{1}$ is determined in terms of $A^{0}_{2}$. Equating
equations (3.18) and (3.19), we eliminate~$A_{1}^{0}$ and we get,
\begin{gather}
A^{2}_{1} = - \frac{q^{2}}{\bar{J}} A^{0}_{2}.
\end{gather}

From this, we can substitute $A^{2}_{1}$ into the fifth pair of
equations in (3.16), and we obtain~$A^{2}_{2}$ in terms of~$A^{0}_{2}$,
as
\begin{gather}
A^{2}_{2} = \frac{\bar{J}}{A^{0}_{2}}.
\end{gather}
Using the first column of (3.16), we substitute
the relation $A_{1}^{1} = \partial \ln A^{0}_{2}$ from equation
(3.16-2) as well as (3.19) through (3.21) into
equation (3.16-3). In this way,
we can eliminate the coefficients $A_{1}^{0}$, $A_{1}^{1}$,
$A^{2}_{1}$ and $A^{2}_{2}$ from differential equation (3.16-3).
We obtain a partial differential
equation for the function $A^{0}_{2}$ in the form
\begin{gather}
\bar{\partial} \partial \ln A^{0}_{2} - \frac{q^{2}}{\bar{J}}
\left(A^{0}_{2}\right)^{2} + \frac{|J|^{2} \bar{J}}{q^{2} \left(A_{2}^{0}\right)^{2}}
+ \frac{|J|^{2}}{q^{2}} - q^{2} = 0.
\end{gather}
Moreover, if we introduce a new variable defined by
\begin{gather}
Q=  \frac{q A^{0}_{2}}
{i \bar{J}^{1/2}},
\end{gather}
into (3.22), it is transformed into the following form
\begin{gather}
\bar{\partial} \partial \ln Q - \frac{|J|^{2}}{Q^{2}} + Q^{2} = 0.
\end{gather}
This coincides with equation (3.7). Substituting
(3.19)--(3.21) into the pair
of Riccati equations (3.13), we obtain
\begin{gather}
\partial y = - \frac{|J|^{2}}{q^{2} A^{0}_{2}} +
\partial \ln A^{0}_{2} \, y
- \frac{q^{2} A^{0}_{2}}{\bar{J}}  y^{2},   \qquad  \bar{\partial} J =0, \nonumber\\
    \bar{\partial} y = A^{0}_{2} + \bar{\partial} \ln \left(\frac{\bar{J}}
{q^{2} A^{0}_{2}}\right) y + \frac{\bar{J}}{A^{0}_{2}} y^{2}, \qquad
\partial \bar{J} = 0.
\end{gather}
The compatibility condition for (3.25) is satisfied
identically whenever (3.7) holds.

Note that $Q$ is some solution to (3.7) which is related to
$q$ by (3.23). We could obtain a particular form for $A_{2}^{0}$
by considering the case in which $Q=q$. Then (3.23) implies that
\[
A^{0}_{2} = i \sqrt{\bar{J}}.
\]
Equation (3.7) is invariant under the transformation
\begin{gather}
J \rightarrow \lambda J,
\end{gather}
provided that $|\lambda|^{2} =1$. A B\"{a}cklund
parameter $\lambda$ can be introduced into equations (3.25) by carrying
out transformation~(3.26).
In this case, the pair of Riccati equations (3.25) takes the form
\begin{gather}
\partial y = i \lambda^{1/2} \left(\frac{ J \bar{J}^{1/2}}{q^{2}}
- \displaystyle  \frac{q^{2}}{\bar{J}^{1/2}} y^{2}\right),  \qquad
\bar{\partial} J = 0,\nonumber\\
\bar{\partial} y = i \bar{\lambda}^{1/2} \bar{J}^{1/2} + \bar{\partial} \ln
\left(\frac{\bar{J}^{1/2}}{q^{2}}\right) y -i\bar{\lambda}^{1/2} \bar{J}^{1/2} y^{2},
\qquad  |\lambda|^{2} = 1,
\end{gather}
where $q$ satisfies (3.7).
Hence the above DCs become an Auto-BT for GW
system (3.7), while the Darboux transformation is defined by (3.12).
Furthermore, by linearizing the Riccati system (3.27),
we obtain the associated linear spectral problem for (3.7) with spectral
parameter $\mu$, of the form
\begin{gather}
\partial \left(
\begin{array}{c}
\phi_{1}  \\
\phi_{2}
\end{array}   \right) =
\left(
\begin{array}{cc}
0  &  i \mu^{1/2}  \displaystyle \frac{\bar{J}^{1/2} J}{q^{2}}  \vspace{1mm}\\
i q^{2} \displaystyle \left(\frac{\mu}{\bar{J}}\right)^{1/2}   &  0
\end{array}   \right)
\left(
\begin{array}{c}
\phi_{1}  \\
\phi_{2}
\end{array}  \right),
\nonumber\\
\bar{\partial}  \left(
\begin{array}{c}
\phi_{1}  \\
\phi_{2}
\end{array}   \right) =
\left(
\begin{array}{cc}
\displaystyle \frac{1}{2} \bar{\partial} \ln (\frac{\bar{J}^{1/2}}{q^{2}})  & i\bar{\mu}^{1/2} \bar{J}^{1/2}  \vspace{1mm}\\
i \bar{\mu}^{1/2} \bar{J}^{1/2}    &  \displaystyle -\frac{1}{2} \bar{\partial} \ln (\frac{\bar{J}^{1/2}}
{q^{2}})
\end{array}    \right)
\left(
\begin{array}{c}
\phi_{1}  \\
\phi_{2}
\end{array}  \right),
\end{gather}
where $y= \phi_{1}/ \phi_{2}$, $|\mu|^{2}=1$ and $\bar{\partial} J= 0$.
The Lax pair (3.28) is based on a nondegenerate $sl(2, \mathbb C)$
representation.
Note that for any holomorphic
function $J$, the compatibility condition for (3.28) reproduces
the system (3.7) in the variable $q$.

\section{Multi-soliton solutions}

A number of new types of soliton solutions to the GW system (3.1) will
now be presented here based on the BT (3.27).
Let us define a new dependent variable
\begin{gather}
\rho = \frac{\psi_{1}}{\bar{\psi}_{2}}.
\end{gather}
It has been shown in [23], that if functions $\psi_{1}$ and $\psi_{2}$
are solutions of GW system (3.1), then the function $\rho$
defined by (4.1) is a solution of the two-dimensional Euclidean
sigma model equations
\begin{gather}
\partial \bar{\partial} \rho - \frac{2 \bar{\rho}}{1 + |\rho|^{2}}
\partial \rho \bar{\partial} \rho = 0, \qquad
\partial \bar{\partial} \bar{\rho} - \frac{2 \rho}{1 + |\rho|^{2}}
\partial \bar{\rho} \bar{\partial} \bar{\rho} = 0.
\end{gather}
Conversely, if $\rho$ is a solution of the sigma model (4.2), then
the solutions $\psi_{1}$ and $\psi_{2}$ of GW system (3.1) have the
form
\begin{gather}
\psi_{1} = \epsilon \rho \frac{(\bar{\partial} \bar{\rho})^{1/2}}
{1 + |\rho|^{2}}, \qquad
\psi_{2} = \epsilon \frac{(\partial \rho)^{1/2}}{1 + |\rho|^{2}},
\qquad \epsilon = \pm 1.
\end{gather}
Thus solutions to GW system (3.1) can be obtained directly by
applying the transformation~(4.3) when a solution of the
sigma model (4.2) is known. Once we have found particular solutions
$\hat{\psi}_{i}$, we can calculate the corresponding value
for the function $q= |\hat{\psi}_{1}|^{2} + |\hat{\psi}_{2}|^{2}$.
Next, we employ succesively the Auto-BT (3.27) in order to find new
solutions $p$ and subsequently from (3.1), we construct the
multi-soliton solution $\psi_{i}$ for GW system (3.1).

1. First, we look for simple
nonsplitting rational soliton solutions of (4.2)
admitting one simple pole given by~[27]
\[
\rho_{j} = \frac{z - a_{j}}{\bar{z} - \bar{a}_{j}},
\qquad j=1, \ldots , N  ,  \qquad a_{j} \in \mathbb C.
\]
Using the Auto-BT (3.27), we get the following algebraic
multi-soliton solution of (3.1)
\begin{gather*}
\psi_{1} = \frac{\epsilon}{2} \left(\sum_{j=1}^{N}
\frac{1}{(\bar{z} - \bar{a}_{j})} \prod_{k=1}^{N}
\frac{z - a_{k}}{\bar{z} - \bar{a}_{k}}\right)^{1/2},
\nonumber\\
\psi_{2} = \frac{\epsilon}{2} \left(\sum_{j=1}^{N}
\frac{1}{(z- a_{j})} \prod_{k=1}^{N}
\frac{z -a_{k}}{\bar{z} - \bar{a}_{k}}\right)^{1/2}, \qquad
\epsilon= \pm 1.
\end{gather*}
For $N=1$, the surface is determined by the equation
\begin{gather}
X_{3} = \frac{1}{2} \ln \frac{4\left(X_{1}^2 + X_{2}^2 -1\right)}
{a_{1}^2 \left(X_{1}^2 +(X_{2}-1)^2\right)}.
\end{gather}
The corresponding surface
of revolution with constant mean curvature $H=1$
for $a_{1}=2$ is plotted in Fig.~1.
Such a surface in $\mathbb R^3$ with a
similar shape has recently been obtained in a cosmological
application to white hole fissioning~[28].

\begin{figure}[th]

\centerline{\epsfig{file=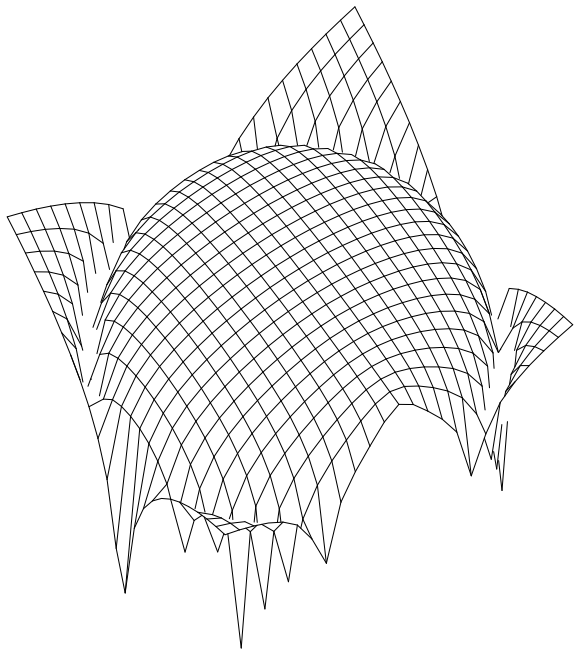,width=0.5\linewidth}}

\caption{Constant mean curvature surface corresponding to equation
(4.4) for $a_{1} = 2$.}
\end{figure}

2. A large class of hyperbolic nonsplitting solutions,
($\partial \bar{\partial} \rho \neq 0$),
of the sigma model equations (4.2) can be constructed when
the function $\rho$ satisfies the algebraic
constraint $|\rho|^{2} = 1$.
Consider a class of nonsplitting hyperbolic solutions of (4.2)
\[
\rho = \sum_{i=1}^{N} \exp(\cosh(z-a_{i}) - \cosh( \bar{z} - \bar{a}_{i})).
\]
In all of the solutions of (4.2)
which are presented below, the $a_{i}$ will be arbitrary complex constants.
The derivatives of $\rho$ with respect to $\partial$ and $\bar{\partial}$
are given by
\begin{gather}
\partial \rho = \sum_{i=1}^{N} \sinh( z- a_{i}) \rho,
\qquad
\bar{\partial} \rho = - \sum_{i=1}^{N} \sinh( \bar{z} - \bar{a}_{i}) \rho.
\end{gather}
respectively. Substituting (4.5)
into (4.3), we obtain the following solutions to GW system~(3.1),
\begin{gather}
\psi_{1} = \frac{\epsilon}{2} \left( \bar{\rho} \sum_{i=1}^{N}
\sinh( \bar{z} - \bar{a}_{i})\right)^{1/2}, \qquad
\psi_{2} = \frac{\epsilon}{2} \left(\rho \sum_{i=1}^{N}
\sinh( z - a_{i})\right)^{1/2},   \nonumber\\
p= \frac{1}{2} \left|\sum_{i=1}^{N} \sinh (z - a_{i}) \right|.
\end{gather}
Note that these solutions do not admit any singularities.
Now, solution (4.6) for $N=1$ is substituted into integrals
(3.3), we obtain
\begin{gather}
X_{1} + i X_{2} = i \sinh(-\cosh(z-a) + \cosh( \bar{z} - \bar{a})),
\nonumber\\
X_{1} - i X_{2} = -i \sinh(-\cosh(\bar{z}- \bar{a})+ \cosh(z-a)),
\nonumber\\
X_{3} = - \cosh(\cosh(z-a) - \cosh( \bar{z} - \bar{a})).
\end{gather}
Eliminating the $z$-dependent factors on the right of (4.7), the following
relationship between the $X_{i}$ variables is obtained
\begin{gather}
X_{1}^{2}+ X_{2}^{2}+ X_{3}^{2} = 1.
\end{gather}
This represents a sphere of unit radius.
Note that similar results hold when $\sinh$ is used in place of $\cosh$ in
expression~(4.4).

3. Consider a class of hyperbolic nonsplitting solution of (4.2)
which are obtained from the $\tanh$ function that satisfies the
algebraic condition $|\rho|^{2} =1$. This type of solution
represents kink-type solution, and is generated by
\begin{gather}
\rho = \sum_{i=1}^{N} \exp( \tanh (z- a_{i}) - \tanh ( \bar{z} -
\bar{a}_{i})).
\end{gather}
The derivatives of $\rho$ in (4.9) are
\begin{gather}
\partial \rho = \sum_{i=1}^{N} {\rm sech}^{2} (z - a_{i}) \rho,
\qquad
\bar{\partial} \rho =- \sum_{i=1}^{N} {\rm sech}^{2} (\bar{z} - \bar{a}_{i})
\rho.
\end{gather}
Substituting (4.10) into (4.3), we obtain the
following multi-soliton solutions $\psi_{i}$
to GW system~(3.1),
\begin{gather}
\psi_{1} = \frac{\epsilon}{2}\left( \bar{\rho} \sum_{i=1}^{N}
{\rm sech}^{2} (\bar{z} - \bar{a}_{i})\right)^{1/2},  \qquad
\psi_{2} = \frac{\epsilon}{2} \left(\rho \sum_{i=1}^{N}
{\rm sech}^{2} (z - a_{i})\right)^{1/2},   \nonumber\\
p = \frac{1}{2} \left|\sum_{i=1}^{N} {\rm sech}^{2} (z - a_{i})\right|.
\end{gather}
Note that the functions $\psi_{i}$ admit only simple poles. For
$N=1$ the surface represents a~sphere of radius one.

4. All functions $\rho$ which generate solutions of (4.2)
in Examples 1 and 2 can be used to generate larger classes
of solution by taking the functions which appear in the
sums for~$\rho$ in expressions (4.4) and (4.9) and combining them by
taking products in different ways.
For example, there exists a hyperbolic, nonsplitting
solution of (4.2) of the form
\begin{gather}
\rho = \exp\left( \sum_{i=1}^{N} ( \cosh( z- a_{i}) - \cosh(
\bar{z} - \bar{a}_{i}) + \sinh( z- a_{i}) - \sinh( \bar{z}
- \bar{a}_{i})\right).
\end{gather}
The corresponding solution of GW system (3.1) has the following form
\begin{gather}
\psi_{1}= \frac{\epsilon}{2} \left( \bar{\rho} \sum_{i=1}^{N}
(\sinh( \bar{z} - \bar{a}_{i}) + \cosh( \bar{z} - \bar{a}_{i}))\right)^{1/2},
\nonumber\\
\psi_{2} = \frac{\epsilon}{2} \left( \rho \sum_{i=1}^{N}
( \sinh( z- a_{i}) + \cosh( z- a_{i}))\right)^{1/2},
\nonumber\\
p = \frac{1}{2} \left| \sum_{i=1}^{N} (\sinh (z - a_{i}) +
\cosh( z - a_{i}))\right|.
\end{gather}
This solution does not admit any singularities.
For $N=1$, the surface represents a sphere of radius one exactly
of the form (4.8).

5. Let the complex function
$g_{i} (z, \bar{z})$ be a set of $N$ harmonic functions
and $f_{i} (z)$ a~set of $N$ arbitrary complex valued functions of
one complex variable $z$, and the complex conjugates of these.
Then we can write a general solution of the sigma model as follows
\[
\rho = \exp\left[ -i \sum_{j=1}^{N} g_{j} (z, \bar{z})\right]
\prod_{j=1}^{N} \frac{f_{j} (z)}{\bar{f}_{j} (\bar{z})}.
\]
Making use of transformation (4.3), the corresponding
general solution of GW system~(3.1) has the form
\begin{gather}
\psi_{1} = \frac{\epsilon}{2}\left[ \rho \left[ i \sum_{j=1}^{N}
\bar{\partial} g_{j} + \sum_{j=1}^{N}
\frac{\bar{\partial} \bar{f}_{j} (\bar{z})}{\bar{f}_{j} (\bar{z})}\right]\right]^{1/2},
\nonumber\\
\psi_{2} = \frac{\epsilon}{2} \left[\rho \left[-i \sum_{j=1}^{N} \partial g_{j}
+ \sum_{j=1}^{N} \frac{\partial f_{j} (z)}{f_{j} (z)}\right]\right]^{1/2},
\qquad \epsilon = \pm 1.
\end{gather}

6. As a final example, let us consider a
trigonometric solution generated by $\rho$
of the form
\[
\rho = \lambda \sin ( n z),   \qquad n \in \mathbb Z, \qquad \lambda \in
\mathbb C.
\]
Using (4.3), the corresponding periodic solution of GW system (3.1)
has the form
\[
\psi_{1} = \epsilon \lambda \sin (nz)
\frac{(n \bar{\lambda} \cos (n \bar{z}))^{1/2}}
{1 + |\lambda|^{2} \sin(n z) \sin(n \bar{z})},
\qquad
\psi_{2} = \epsilon
\frac{(n \lambda \cos(nz))^{1/2}}
{1 + |\lambda|^{2} \sin(nz) \sin (n \bar{z})}.
\]
The corresponding constant mean curvature surface for $\lambda=1$ is given by
\[
X_{1}^{2} + X_{2}^{2} =
\frac{X_{3} (X_{3}-2)^{2}}{4-X_{3}},
\]
which has the shape of a `Mexican hat'. It is obtained by rotation
of the curve
\[
X_{1}^2 (4 - X_{3}) - X_{3} (X_{3} - 2)^2 =0,
\]
for $X_{3} \in [0,4)$, around the $X_{1}$ or $X_{2}$ axis.

It is useful at this point to mention some physical
applications of some of these surfaces.
Cylinders and spheres have applications to certain
types of cosmological models, and should also be useful
in describing event horizons in general relativity~[1].
In string theory, the motion of a particle is described
by a surface which propagates through space-time. Minimal
surfaces can constitute a way of describing particle states.

Certain types of soliton solutions which are localized
in time as well as in space are referred to as instantons.
Such solutions exist in gauge theories, since the gauge-field
equations are relativistic and give rise to topological
nontriviality in time as well as in space~[29]. The application
of the generalized Weierstrass system to strings in
three-dimensional Euclidean space will be outlined.
This system allows the construction of any surface in $\mathbb R^3$,
where $p$ is a real-valued function of $z$ and $\bar{z}$.
With $u = |\psi_{1}|^2 + |\psi_{2}|^2$, the Gaussian curvature
is $K=- u^{-2} \partial \bar{\partial}  (\log u)$ and the mean curvature
is $H=p/u$, so that when the mean curvature is constant, $p=u$.
In terms of the variables $p$, $\psi_{i}$ the required action
for the string has the following form
\[
S = 4 \mu_{0} \int \left(|\psi_{1} |^2 + |\psi_{2}|^2\right)^2 \,
dx dy + \frac{4}{\alpha_{0}} \int p^2 \, dx dy.
\]
Classical configurations of strings can be described
by common solutions of this Nambu--Goto--Polyakov action $S$
and the generalized Weierstrass equations, which provide
surfaces in $\mathbb R^3$. In generic coordinates, the
corresponding Euler--Lagrange equation has the form
\begin{gather}
\Delta H + 2 H \left(H^2 -K\right) - 2 \alpha_{0} \mu_{0} H = 0,
\end{gather}
where $\Delta$ is the Laplace--Beltrami operator, and under the
conformal metric
\[
\Delta H = u^{-2} \partial \bar{\partial} H.
\]
In terms of the variable $\varphi = H^{-1}$, $p=u/\varphi$,
the Euler--Lagrange equation takes the form
\[
\partial \bar{\partial} \varphi
+ \left[2 p^2 + \partial \bar{\partial} \ln p^2\right] \varphi
- 2 \alpha_{0} \mu_{0} p^2 \varphi^3 = 0.
\]
When the mean curvature is constant, $\varphi = \varphi_{0}$,
and the Euler--Lagrange equation reduces to a second order
linear equation
\[
\partial \bar{\partial} \ln p^2 + 2 \left(1 - \alpha_{0} \mu_{0}
\varphi_{0}^2\right) p^2 = 0,
\]
which can be transformed into the Liouville equation
\begin{gather}
\partial \bar{\partial} \theta + \beta e^{\theta} = 0,
\end{gather}
where $\theta = \ln p^2$. For $\beta \neq 0$, the solution of
(4.15) has the form
\[
\theta = \ln \partial G + \ln \bar{\partial} \bar{G} -
2 \ln \left(|G|^{2} + \beta/2\right),
\]
where $G(z)$ is
an arbitrary analytic function.
Solutions of (4.14) which correspond to mean curvature different
from zero can be invesigated as well. This task will be
undertaken in a future work.

\section{Final remarks}

In this paper, we demonstrate that the
task of finding large classes of solutions of PDEs is related
to the group properties of an overdetermined system composed of
an initial system of PDEs (2.1) subjected to DCs (2.2). We show that this has
a group theoretical interpretation in terms of ``conditional
symmetries''. The main difficulty in this approach is related to
finding reasonable ansatzes that yield compatible solutions.
The approach adopted here is based on
multiple constraints satisfying several specific conditions,
such as~(2.10), (2.12), (2.13) and (2.14), which enables us to
overcome these difficulties. The proposed approach simplifies the
task of solving the nonlinear determining equations (2.13) for the
initial system (2.9) by providing us with an almost entirely
algorithmic procedure. The most important advantage of this method is
that it gives us effective tools for constructing
certain classes of BT described by first order differential
equations in a~systematic way. Its effectiveness has been
demonstrated by the results, both reconstructed [10, 21, 24, 25] and new
in Section 3 and in Section 4 for the GW system (3.1).

In conclusion, we have learned from this example that it is useful
to subject system~(2.9) with another one, which involves certain
auxiliary variables dictated by the possible representations of the
``hidden'' symmetry algebra (2.12). Selection of appropriate algebras
and their representations is still an open problem, requiring further
investigation. Finally, the presented example of GW system~(3.1)
suggests that there may exist a connection between the conditional
symmetry method and the prolongation structure approach of Wahlquist
and Estabrook~[30]. Indeed, the auxiliary function $y$ is
nothing else than the pseudopotential introduced in their approach.

\subsection{Acknowledgments}

This work was supported by a research grant from NSERC of Canada
and the Fonds FCAR du Gouvernment du Qu\'{e}bec.
We would like to thank Professor R. Conte (CEA Saclay) for
helpful and interesting discussions on the topic of the paper.

\label{bracken-lastpage}

\end{document}